\documentclass[
  prd,
  aps,
  showpacs,
  eqsecnum,
  notitlepage,
  nofootinbib,
  superscriptaddress,   
  colorlinks=true,
  linkcolor=blue,
  citecolor=blue,
  urlcolor=blue,
  pdfstartview=FitV,
  breaklinks=true,
  10pt,twocolumn
]{revtex4-2}

\usepackage[utf8]{inputenc}
\usepackage[T1]{fontenc}
\usepackage{lmodern}
\usepackage{slashed}
\usepackage{bbm}
\usepackage{xcolor}

\usepackage{graphicx}
\usepackage{tikz}
\usepackage{siunitx}
\usepackage{tensor}
\usepackage{amsmath,amssymb,amsfonts,bm}
\usepackage{slashed}
\usepackage{enumerate}
\usepackage{scalerel}       
\usepackage{accents}
\usepackage{xfrac}

\usepackage[caption=false]{subfig}

\usepackage{orcidlink}
\begin{document}

\title{Nonlocality effects on fermions confined within a hard-wall potential}

\author{G.~Q.~Garcia\,\orcidlink{0000-0003-3562-0317}}
\email{gqgarcia99@gmail.com}
\affiliation{Centro de Ciências, Tecnologia e Saúde, Universidade Estadual da Paraíba, 58233-000, Araruna, PB, Brazil}
\author{P.J. Porf\'irio\,\orcidlink{0000-0002-4713-3332}}
\email{pporfirio@fisica.ufpb.br}
\affiliation{Departamento de F\'isica, Universidade Federal da Para\'iba, 58051-970, Jo\~ao Pessoa, PB, Brazil. }
\author{E. Brito\, \orcidlink{0000-0002-6967-2126}}
\email{eliasbaj@ufob.edu.br}
\affiliation{Departamento de F\'isica, Universidade Federal da Para\'iba, 58051-970, Jo\~ao Pessoa, PB, Brazil. }
\affiliation{Centro de Ci\^encias Exatas e das Tecnologias, Universidade Federal do Oeste da Bahia, 47810-059, Barreiras, BA, Brazil}
\author{D. C. Moreira\,\orcidlink{0000-0002-8799-3206}}
\email{moreira.dancesar@gmail.com}
\affiliation{Centro de Ciências, Tecnologia e Saúde, Universidade Estadual da Paraíba, 58233-000, Araruna, PB, Brazil}

\begin{abstract}
 We investigate nonlocality effects on the energy spectrum of spin-$1/2$
fermions confined by hard-wall potentials. Starting from a nonlocal Dirac
theory with a modified kinetic term, we formulate the associatd field equation
in flat spacetime using cylindrical coordinates. By exploiting the eigenvalue
spectrum of the squared Dirac operator, the original infinite-order  equation
is reduced to a finite-order problem. We also show that on the branch continuously
connected to the local Dirac theory the nonlocal corrections can be encoded
in an effective fermion mass. We then find analytical energy
spectra for two confining geometries: a cylindrical hard wall and a quantum
ring defined by two concentric cylindrical boundaries. In both cases, the
radial equations reduce to Bessel equations and the boundary conditions
quantize the allowed modes. The local hard-wall spectra are recovered for
$\Lambda\to\infty$, whereas finite nonlocality shifts the mass contribution and, consequently,  the particle--antiparticle energy separation. These results
provide a   direct characterization of nonlocal effects in quantum-confined
fermionic systems.
\end{abstract}


\maketitle
\section{Introduction}
\label{intro}

Recent advances in the study of spin-1/2 particle dynamics in curved backgrounds and topological defect spacetimes have revealed promising prospects for the development of novel semiconductor devices. This framework is particularly relevant for the description of nanostructured systems, such as quantum rings, quantum dots, and graphene quantum dots, where quantum confinement, surface curvature, and external fields play a fundamental role in determining their optical, electrical, and magnetic properties \cite{bajar2025non,carvalho2026new}. Roostaei demonstrated that one-dimensional quantum rings preserve the magnitude of the electric dipole moment while allowing fluctuations in its orientation. When multiple quantum rings are arranged in an array, the system exhibits antiferroelectric ordering provided that the charge distribution remains discrete. However, this ordering disappears in the continuum limit. Furthermore, in the presence of external magnetic fields at zero temperature, the system undergoes a quantum phase transition accompanied by shifts in the energy eigenstates \cite{roostaei2008polarization}. Experimental studies have also shown that GaSb/GaAs quantum dots may spontaneously evolve into quantum ring structures during the growth process \cite{timm2008self}. From an optical perspective, double-ring quantum dot systems exhibit a pronounced blueshift in their emission spectra \cite{bietti2009quantum}. Likewise, graphene-based nanostructures display remarkable electronic and optical properties arising from quantum confinement and edge effects, which induce energy-gap opening and generate photoluminescent responses highly sensitive to both geometry and system size \cite{li2015electronic}.

The theoretical foundations underlying these phenomena can be traced back to the pioneering works of Katanaev \cite{katanaev1992theory} and Landau \cite{landau1930diamagnetismus}, who established a geometric framework for describing topological defects in condensed matter systems. Within the Einstein–Cartan–Riemann formalism, curvature and lattice defects are treated through the language of differential geometry, allowing crystalline defects to be represented as effective geometric structures \cite{harvey2012einstein}. In this approach, particle motion is governed by geodesic equations, providing a natural description of quantum dynamics on curved surfaces and defected media. This geometric formulation has become a powerful tool for investigating the interplay between topology, curvature, and quantum mechanics in low-dimensional systems\cite{trautman2006einstein, ruggiero2003einstein,da1981quantum}.

One of the most important consequences of this framework is the sensitivity of quantum energy spectra to external and geometric conditions. Significant modifications of the energy levels arise from changes in the geometry of the confining surface, the presence of topological defects, the application of nonlinear or oscillating magnetic fields, and confinement mechanisms described by hard-wall potentials \cite{bakke2013noninertial}. Since geometry directly affects the spectral structure of the system, several phenomena exhibit analogies with well-known quantum and geometric effects. For instance, disclination-type defects in quantum rings generate effective contributions analogous to those associated with the Aharonov–Bohm effect \cite{azevedo1998topological}, while the Aharonov–Bohm flux may compensate for quantum effects induced by screw dislocations in crystalline media \cite{da2021effects}.

Using the geometric theory of defects, Silva Netto and collaborators investigated the dynamics of quantum particles confined in one- and two-dimensional quantum rings. Their results showed that the Aharonov–Bohm flux can act as an experimental fine-tuning mechanism capable of compensating for energy-level shifts produced by topological defects, while hard-wall potentials provide an accurate description of particle confinement within ring-shaped nanostructures \cite{netto2008influence}. Moreover, noninertial effects were shown to induce an effective hard-wall confinement for neutral spin-1/2 particles, leading to the formation of bound states. External field configurations may also generate additional bound states, enriching the spectral structure of the system.

Further investigations by Silva and Bakke examined topological effects associated with spiral dislocations in quantum rings. Their analysis revealed that the influence of such defects manifests through the emergence of an effective radius in the system, whereas Aharonov–Bohm-type effects for bound states do not arise in this class of dislocations \cite{da2020topological}. Subsequently, Belich and Bakke studied the influence of nonuniform magnetic fields in the same geometric background, introducing a Lorentz-symmetry-violating scenario that affects the energy spectrum without modifying the topological contributions associated with persistent spin currents and rotational effects. Interestingly, although Lorentz-symmetry violation alters the energy levels of the system, the persistent currents induced by geometric and topological phases remain unaffected \cite{bakke2025aharonov}. These findings reinforce the central role played by geometry, topology, and external fields in controlling quantum phenomena in low-dimensional systems and its applications.

The study of nonlocal theories has attracted considerable attention in high-energy physics \cite{wataghin1934bemerkung,wataghin1953non}. Owing to their ability to regulate the ultraviolet (UV) sector and prevent the emergence of divergent quantities, nonlocal frameworks have become valuable tools for investigating interactions at the Planck scale. Such theories have been employed, for instance, in the analysis of the Higgs mechanism within quantum field theories \cite{hashi2018higgs} and in one-loop corrections to generalized scalar quantum electrodynamics (QED) models \cite{gama2018spontaneous}. In 1967, Efimov proposed a nonlocal theory based on the construction of a perturbative scattering matrix free from ultraviolet divergences, illustrating the formalism through the quantization of a scalar field. As a consequence, at sufficiently large distance scales, higher-order perturbative contributions introduce small but finite violations of causality \cite{efimov1967non}. On the other hand, nonlocal theories have also found important applications in the low-energy regime. This perspective emerged naturally from string and M-theory frameworks, where nonlocal operators arise as effective descriptions of low-energy phenomena \cite{witten1986non}. When applied to gravity interactions, nonlocal theories are commonly regarded as part of the broader class of Extended Theories of Gravity (ETGs) \cite{capozziello2023energy}. In this context, two major classes of nonlocal gravitational models are usually identified: Infinite Derivative Gravity (IDG) and Integral Kernel Gravity (IKG). In IDG, the nonlocal operator is typically expressed as $f(\Box/\Lambda^2)$, where $f$ is an entire analytic function, $\Box$ denotes the d'Alembert operator, and $\Lambda$ characterizes the fundamental nonlocality scale.

In 2007, Deser and Woodard developed a nonlocal gravitational theory where the nonlocal operator emerges from quantum loop corrections \cite{deser2007nonlocal}. Their model provided a mechanism capable of explaining the current accelerated expansion of the universe without invoking a dark-energy component. The construction relies on a retarded Green's function; however, the resulting solution preserves the same sign both in gravitationally bound systems and in an expanding cosmological background. In 2014, the authors refined the formalism and derived a nonlocal operator whose sign changes appropriately between these two physical regimes \cite{deser2019nonlocal}. Recently, Bajardi and Capozziello \cite{bajardi2022non} investigated nonlocal operators of the form $\Box^{-1}R$, where $R$ denotes the Ricci scalar. By selecting a specific functional form and using the Noether symmetry approach in spherically symmetric spacetimes, they compared the resulting predictions with observational data from the orbit of the S2 star. Similarly, Acunzo and collaborators considered theories described by nonlocal operators of the form $f(R,\Box^{-1}R)$ and applied the Noether symmetry method to identify viable solutions in cosmological minisuperspace models \cite{acunzo2022non}. Their analysis was based on the localization procedure in which the nonlocal quantity $\Box^{-1}R$ is replaced by an auxiliary scalar field $\phi = \Box^{-1}R$, or $R = \Box \phi$, thereby reproducing the effects of nonlocality at the level of the localized Lagrangian. Since the present work focuses on fermionic fields, it is natural to adopt the Dirac operator as the argument of the nonlocal function $f$. Consequently, the nonlocal operator assumes the form $g(\Box,\slashed{\partial})$, providing a natural extension of nonlocal gravitational frameworks to fermionic sectors.

In this work, we investigate the dynamics of spin-1/2 fermions within the framework of nonlocal field theories, following the approach developed in \cite{nascimento2026novel}. We consider fermionic particles confined in a flat spacetime background with cylindrical symmetry, where the nonlocality operator is constructed as a function of the Dirac operator. Specifically, the nonlocal structure is introduced through a functional operator $g(\Box,\slashed{\partial})$, thereby extending conventional local fermionic dynamics to include nonlocal corrections. The underlying geometry is described in cylindrical coordinates, allowing the realization of a quantum-ring configuration and providing a suitable platform for exploring the interplay between geometry, topology, and nonlocality. Our main goal is to determine how nonlocal contributions modify the quantum dynamics of confined fermions and, in particular, how they affect the energy spectrum associated with  cylindrical hard-wall confinement. By solving the corresponding nonlocal Dirac equation for a cylindrical box and a quantum-ring geometry, we analyze the corrections induced by the nonlocal operator on the bound-state energy levels and investigate the resulting spectral shifts. The present study therefore establishes a connection between nonlocal quantum field theories and geometrically confined quantum systems, providing new insights into the role of nonlocal effects in flat spacetime with nontrivial boundary conditions.

This work is organized as follows. In Sec. II, we discuss the key properties of nonlocal fermion theory. In Sec. III, we formulate a nonlocal Dirac equation in cylindrical coordinates, incorporating the Dirac matrices and the nonlocality parameter, and in Sec. IV we solve this equation with an ansatz which decouples the equation for each spinor component. In Sec. V, we  find the energy levels of nonlocal fermions confined in a cylinder with a hard-wall potential. In Sec. VI, we capture the energy eigenvalues for nonlocal fermions confined in a quantum ring. Finally, in Sec. VII, we present our final remarks.

\section{nonlocal fermion theory}\label{nonLocalTheory}

We start here by briefly reviewing some of the main properties of the nonlocal spin-1/2 field theory proposed in \cite{nascimento2026novel}. Before proceeding, it is useful to clarify an important feature of the model. Nonlocality is implemented by introducing an infinite-order differential operator, usually referred as a form factor, into the kinetic term of the Dirac Lagrangian. The form factor is chosen to be a non-polynomial entire function of the Dirac operator. In particular, the exponential form factor adopted below is entire and zero-free and therefore does not introduce additional singularities through the form factor itself, avoiding the presence of undesirable ghost-like degrees of freedom, which spoil the unitarity of the quantum theory. In this scenario, the nonlocal Dirac equation is given by
\begin{equation}
        [i\slashed{\partial} f(\slashed{\partial})-m]\Psi=0.
    \label{eq2}
\end{equation}
By multiplying the left side of the former equation by the operator $i\slashed{\partial} f(\slashed{\partial})+m$, as is usually done for the local case, one finds
\begin{equation}
        [\slashed{\partial}^{2} f^{2}(\slashed{\partial})+m^{2}]\Psi=0.
    \label{klein}
\end{equation}
Note that $\slashed{\partial}^2=-\Box \hat{I}$, where $\Box=\eta^{\mu\nu}\partial_{\mu}\partial_{\nu}$ denotes the d'Alembert operator.  The minus sign appears in front of the d'Alembert operator because we are adopting the mostly-plus metric signature $(-,+,+,+)$. 

An interesting point on this context is that Eq. \eqref{klein} can be rewritten in a more suitable way by noting that any function of the Dirac operator can be expanded as a linear combination of the basis elements $\{\hat{I},\slashed{\partial}\}$. Let us see this in more detail. Since $f(\slashed{\partial})$ is an entire function, one can expand it in a power series of its argument, i.e.,
\begin{eqnarray}
\nonumber f(\slashed{\partial})&=& \left[\sum_{n=0}^{\infty}b_{n}\slashed{\partial}^n_{\Lambda} \right]\\
\nonumber&=&\left[\sum_{n=0}^{\infty}b_{2n}\slashed{\partial}^{2n}_{\Lambda}+\sum_{m=0}^{\infty}b_{2m+1}\slashed{\partial}^{2m+1}_{\Lambda}\right]\\
\nonumber&=&\left[\sum_{n=0}^{\infty}b_{2n}({-1})^{n}\frac{\Box^{n}}{\Lambda^{2n}}\right]\hat{I}+\\
\nonumber &+&\left[\sum_{m=0}^{\infty}b_{2m+1}({-1})^{m}\frac{\Box^m}{\Lambda^{2m+1}}\right]\slashed{\partial}\\
&=& f_{1}(\Box)\hat{I}+f_2(\Box)\slashed{\partial},\label{k}
\end{eqnarray}
where $f_{1}(\Box)$ and $f_{2}(\Box)$ are form factors of the d'Alembert operator and, for simplicity, we define  $\slashed{\partial}_{\Lambda}\equiv\slashed{\partial}/\Lambda$. The mass scale $\Lambda$ is a typical high-energy scale at which the nonlocality effects become more relevant.
From the previous equation, it is straightforward to see that
\begin{equation}
    f^2(\slashed{\partial})=a_{1}(\Box)\hat{I}+a_2(\Box)\slashed{\partial},
    \label{kj}
\end{equation}
with $a_1(\Box)=f^2_1(\Box)-\Box f_2^2(\Box)$ and $a_2(\Box)=2f_1(\Box)f_2(\Box)$. Plugging Eq.\eqref{kj} into Eq.\eqref{klein}, one finds
\begin{equation}
    \left[\Box\ g(\Box,\slashed{\partial}) - m^2\right]\Psi = 0,
    \label{nonLocalDirac}
\end{equation}
where $g(\Box,\slashed{\partial})=f^2(\slashed{\partial})$. 

An additional restriction on the choice of nonlocal form factors arises from the requirement that the kinetic operator be Hermitian, as pointed out in  \cite{Porfirio:2026zga}. Adopting the conventional Dirac inner product and assuming appropriate boundary conditions, the covariant Dirac operator satisfies $\slashed{\nabla}^{\dagger}=-\slashed{\nabla}$, ensuring the formal self-adjointness of $i\slashed{\nabla}$. For a kinetic term of the form $i\slashed{\nabla}f(\slashed{\nabla})$, with $f$ analytic, Hermiticity is preserved when the form factor obeys
\begin{equation}\label{6}
f(\slashed{\nabla})^{\dagger}=f(\slashed{\nabla}),
\end{equation}
or, equivalently, $f(z)^*=f(-z^*)$. Unlike the requirement of analyticity and the absence of zeros, this condition constrains the complex structure of the form factor. Hence, the use of an entire, zero-free function, although sufficient to avoid introducing additional zeros through the form factor, does not by itself guarantee a Hermitian kinetic operator. The reality condition above must also be satisfied to ensure a consistent Hermitian formulation in Lorentzian spacetime.

Given the discussion above, the nonlocal spin-1/2 field equation we have to deal with is an infinite-order partial differential equation, for which there is no generic technique to solve it. In particular, here we adopt the mathematical approach proposed in \cite{nascimento2026novel}, which uses the spectrum of eigenvalues of the square Dirac operator to reduce the derivative order of the nonlocal Dirac equation. To do so, consider the eigenvalue equation
\begin{equation}
    \slashed{\partial}^2\Psi=-\Box\Psi=\lambda^2\Psi,
    \label{hj}
\end{equation}
where $\lambda^2$ denotes the eigenvalue of the square Dirac operator. Generically, it can be checked that
\begin{subequations}
\begin{eqnarray}
\label{2m}\slashed{\partial}^{2m+1}\Psi&=&\lambda^{2m}\slashed{\partial}\Psi,\,\,\,\,\,m \in \mathbb{N},\\
\label{2n}\slashed{\partial}^{2n}\Psi&=&\lambda^{2n}\Psi,\,\,\,\,\,n \in \mathbb{N},
\end{eqnarray}
\end{subequations}
and with the identities above in hands one can make the form factor explicit. Let us take
\begin{equation}
    f(\slashed{\partial})=e^{-i\frac{\slashed{\partial}}{\Lambda}}.
\end{equation}
Expanding the exponential in a power series and using the recurrence formulas \eqref{2m} and \eqref{2n}, we find
\begin{eqnarray}
    \nonumber e^{-i\frac{\slashed{\partial}}{\Lambda}}\Psi&=&\left(\sum
        \limits_{n=0}^{\infty}
        \frac{(-1)^n}{2n!}\frac{\slashed{\partial}^{2n}}{\Lambda^{2n}}+\sum
        \limits_{m=0}^{\infty}
        \frac{(-1)^m}{(2m+1)!}\frac{\slashed{\partial}^{2m+1}}{\Lambda^{2m+1}}\right)\!\Psi\\
        &=&\left(\cos\left(\frac{\lambda}{\Lambda}\right)\hat{I}-\frac{i}{\lambda}\sin\left(\frac{\lambda}{\Lambda}\right)\slashed{\partial}\right)\Psi,
\end{eqnarray}
and, as a consequence, 
\begin{equation}
g(\Box,\slashed{\partial})\Psi=e^{-2i\frac{\slashed{\partial}}{\Lambda}}\Psi=\!\left(\!\cos\left(\frac{2\lambda}{\Lambda}\right)\!\hat{I}\!-\!\frac{i}{\lambda}\!\sin\left(\frac{2\lambda}{\Lambda}\right)\!\slashed{\partial}\right)\!\Psi.
\end{equation}
Substituting this equation into Eq. \eqref{nonLocalDirac}, one finds
\begin{equation}
\left[\Box\left(\cos\left(\frac{2\lambda}{\Lambda}\right)\hat{I}-\frac{i}{\lambda}\sin\left(\frac{2\lambda}{\Lambda}\right)\slashed{\partial}\right)-m^2\right]\Psi=0,
\label{kkl}
\end{equation}
which is formally a third-order partial differential equation. Moreover, by using the eigenvalue condition \eqref{hj} together with the zero commutator $[\Box,\slashed{\partial}]=0$, one can show that the Eq. \eqref{kkl} becomes
\begin{equation}
    \left[\cos{\left(\frac{2\lambda}{\Lambda}\right)} \hat{I} - \frac{i}{\lambda} \sin{\left(\frac{2\lambda}{\Lambda}\right)}\ \slashed{\partial}\right] \Psi = -\left(\frac{m}{\lambda}\right)^2\Psi. ~~~~\label{nonLocalDirac3}
\end{equation}
Therefore, the original infinite-order equation is now reduced to a finite-order differential equation on each eigenspace of the squared Dirac operator. Next, we solve the nonlocal Dirac equation found for particular scenarios with confined fermions.
\section{Building the nonlocal Dirac equation in cylindrical coordinates}\label{Building}

In this section we are interested in exploring nonlocal effects arising from the model discussed in the previous section on the energy levels of fermions confined within a hard-wall potential, described by a planar system in cylindrical coordinates. Thus, we must address the nonlocal Dirac equation~\eqref{nonLocalDirac} in order to find analytical expressions for the fermionic energy levels in the related systems.

The next step in confining the fermion field within a hard-wall potential is to place it on a flat spacetime in cylindrical coordinates, which acts as a cylindrical box where the fermion lives. The line element of such a spacetime is
\begin{equation}\label{flatsp}
    ds^2 = -dt^2 + dr^2 + r^2 d\phi^2 + dz^2,
\end{equation}
where $-\infty<t<\infty,~-\infty<z<\infty,~0<\phi<2\pi$ and $r\geq0$ denotes a radial coordinate. In this scenario one can define a non-holonomic basis $\hat{\theta}^a = e^{a}_{\ \mu}(x) dx^{\mu}$ and its respective inverse, so that they satisfy the identity $ds^2 = \eta_{ab}\hat{\theta}^{a}\wedge\hat{\theta}^{b}$. In this way, the {\it vielbien} and its inverse are
\begin{eqnarray}
    e^{a}_{\ \mu}(x) = \begin{bmatrix}
        1 & 0 & 0 & 0\\
        0 & 1 & 0 & 0\\
        0 & 0 & r & 0\\
        0 & 0 & 0 & 1
    \end{bmatrix}~\text{and}~ e_{a}^{\ \mu}(x) = \begin{bmatrix}
        1 & 0 & 0 & 0\\
        0 & 1 & 0 & 0\\
        0 & 0 & \frac{1}{r} & 0\\
        0 & 0 & 0 & 1
    \end{bmatrix},~~
\end{eqnarray}
respectively. The subsequent procedure is to solve the Maurer-Cartan structures with zero torsion, given by $d\hat{\theta}^a + \omega^{a}_{\ b}\wedge\hat{\theta}^{b} = 0$, where the 2-form connection $\omega^{a}_{\ b} = \omega^{\ a}_{\mu \ b}\ dx^{\mu}$ are calculated from spin connections $\omega^{\ a}_{\mu \ b}$. For the flat spacetime \eqref{flatsp}, in particular, the only nonzero spin connetions are  $\omega^{\ 1}_{\phi\ 2} = -\omega^{\ 2}_{\phi\ 1} = -1$. In agreement with the spinor theory~\cite{nakahara2018geometry, weinberg1998gravitation}, we also introduce the covariant derivative $\nabla_\mu = \partial_\mu + \Gamma_\mu$, where  $\Gamma_{\mu} = \frac{i}{4} \omega_{\mu ab}(x) \Sigma^{ab}$ represents the spinorial connection and $\Sigma^{ab} = \tfrac{i}{2}\left[\gamma^a,\gamma^b\right]$, where $\gamma^a$ denotes the Dirac matrices in the representation
\begin{equation}
    \gamma^0 = \begin{pmatrix}
     \mathbbm{1}_{2\times 2} & 0\\
     0 & \mathbbm{1}_{2\times 2}
    \end{pmatrix}\ \ \text{and}\ \  \gamma^i = \begin{pmatrix}
     0 & \sigma^i\\
     \sigma^i & 0
    \end{pmatrix},
\end{equation}
for $i = 1, 2, 3$, and $\{\sigma^i\}$ is the set of Pauli matrices\footnote{It should be remarked that the nonlocal Dirac equation was developed in flat spacetime \cite{nascimento2026novel}. For a curved spacetime, the nonlocal Dirac equation takes a different form of Eq.\eqref{nonLocalDirac}. In our case, its sufficient to promote the partial derivative to a covariant derivative, since the Riemann tensor is zero. Nonlocal Dirac equation in a curved spacetime is treated elsewhere \cite{Porfirio:2026zga}.}.

With all the ingredients described above, the nonlocal Dirac equation in Eq. \eqref{nonLocalDirac3} in cylindrical coordinates is now given by
\begin{equation}
    i \slashed{\mathcal{D}}_{\rm cyl}\Psi=M_{\rm eff}\Psi,
    \label{effective_dirac}
\end{equation}
written in the standard first-order form by defining the cylindrical Dirac operator as
 \begin{equation}
 \slashed{\mathcal{D}}_{\rm cyl}\equiv
\gamma^0 \partial_t
+\gamma^1\left(\partial_r+\frac{1}{2r}\right)
+\frac{\gamma^2}{r}\partial_\phi
+\gamma^3\partial_z,
\label{Dcyl}
\end{equation}
and a dimensionful effective mass
\begin{equation}
     M_{\rm eff}(\lambda,\Lambda)=
    \frac{\lambda}{\sin\!\left(2\lambda/\Lambda\right)}
    \left[\frac{m^2}{\lambda^2}+\cos\!\left(\frac{2\lambda}{\Lambda}\right)\right].
    \label{efffM}
\end{equation}
Thus, the nonlocal contribution is absorbed into the mass sector while the spacetime coordinates retain their original physical meaning. It is important to point out here that the parameter $\lambda$ is not independent on shell. The eigenvalue condition~\eqref{hj}, written in cylindrical coordinates, together with Eq.~\eqref{effective_dirac}, lead us to the identity
\begin{equation}
    \lambda^2=-M_{\rm eff}^{\,2}.
    \label{lambda_meff_relation}
\end{equation}
Indeed, the same result can be directly implemented on the original nonlocal Dirac equation~\eqref{eq2}. Since Eq.~\eqref{effective_dirac} implies $ \slashed{\mathcal{D}}_{\rm cyl}\Psi=-iM_{\rm eff}\Psi$, the exponential form factor acts on an on-shell state as
\begin{equation}
     e^{-i \frac{\slashed{\mathcal{D}}_{\rm cyl}}{\Lambda}}\Psi
    =e^{-\frac{M_{\rm eff}}{\Lambda}}\Psi.
\end{equation}
Consequently, the physical effective mass obeys the transcendental equation
\begin{equation}
     M_{\rm eff}e^{-\frac{M_{\rm eff}}{\Lambda}}=m,
    \label{meff_onshell}
\end{equation}
whose solution is
\begin{equation}
    M^{(n)}_{\rm eff}=-\Lambda W_n\!\left(-\frac{m}{\Lambda}\right),
\end{equation}
where $W_n(z)$ denotes the multivalued Lambert $W$ function, also known as the product logarithm. It possesses infinitely many branches, labeled by $n\in\mathbb{Z}$. Real-valued solutions only arise from the principal branch $W_0(z)$, defined for $z\ge -e^{-1}$, and from the lower branch $W_{-1}(z)$, defined for $-e^{-1}\le z<0$. However, only the principal branch provides the physically relevant solution, since it is regular at $z=0$ and is the unique branch continuously connected to the local theory. In particular, it yields $M_{\rm eff}\to m$ in the limit $\Lambda\to\infty$. Thus,
 the branch continuously connected to the local Dirac theory is
\begin{equation}
     M_{\rm eff}=-\Lambda W_0\!\left(-\frac{m}{\Lambda}\right).
    \label{meff_lambert}
\end{equation}
 From now on, we consider only this branch, while the other branches are disregarded. 
It is noteworthy that in the relevant regime  for the local limit, $m/\Lambda\ll1$, Eq.~ \eqref{meff_lambert} yields
\begin{equation}
     M_{\rm eff}
    =m+\frac{m^2}{\Lambda}
    +\frac{3m^3}{2\Lambda^2}
    +\frac{8m^4}{3\Lambda^3}
    +\mathcal{O}(\Lambda^{-4}).
    \label{meff_expansion}
\end{equation}
Hence $M_{\rm eff}\to m$ smoothly as $\Lambda\to\infty$. A direct consequence of   Eqs.~\eqref{efffM} and \eqref{meff_expansion} is that, on the local branch, nonlocality modifies the mass contribution to the fermionic spectrum. We now use Eq.~\eqref{effective_dirac} to determine the bound states generated by hard-wall boundary conditions.

\section{Solving the nonlocal Dirac equation}

In order to solve the modified Dirac equation \eqref{effective_dirac}, we first decouple the set of differential equations associated to each spinor component. To this end, we use the following   \textit{ansatz} for the fermion wave function:
 \begin{equation}
     \Psi(t,r,\phi,z)=e^{-iEt+ij\phi+ikz}
    \begin{pmatrix}
        \eta_1\\
        \eta_2
    \end{pmatrix},
    \label{spinorAnsatz}
\end{equation}
with $\eta_{1,2}=\eta_{1,2}(r)$. By introducing the above expression into   Eq.~\eqref{effective_dirac}, we find two coupled first-order differential equations,
 {\small
\begin{subequations}
     \begin{eqnarray}
         (E\!-\!M_{\rm eff})\eta_1 &=& \! -i\left[\sigma_x\left(\frac{\partial}{\partial r}\!+\!\frac{1}{2r}\right)\!+i\frac{j}{r}\sigma_y+ik\sigma_z\right]\eta_2,~~~~~~\label{EQ1}\\[4pt]
         (E\!+\!M_{\rm eff})\eta_2 &=& \! -i\left[\sigma_x\left(\frac{\partial}{\partial r}\!+\!\frac{1}{2r}\right)\!+i\frac{j}{r}\sigma_y+ik\sigma_z\right]\eta_1.~~~~~~\label{EQ2}
    \end{eqnarray}\end{subequations}}%
By decoupling the equations above we are led to a pair of second-order differential equations with radial structure given by
\begin{equation}
     r^2\frac{\partial^2\eta_{\ell}}{\partial r^2}
    +r\frac{\partial\eta_{\ell}}{\partial r}
    +\left(r^2\xi^2-\nu^2\right)\eta_{\ell}=0,
    \label{4.3}
\end{equation}
with $\ell=1,2$ and where, for later convenience, we define
\begin{subequations}
     \begin{eqnarray}
         \xi^2 &=&  E^2-k^2-M_{\rm eff}^{\,2},\label{xiDef}\\[4pt]
        \nu^2 &=&  \left(j-\frac{\sigma_z}{2}\right)^2.\label{nuDef}
    \end{eqnarray}\end{subequations}
Here, $\sigma_z=\pm1$ denotes the eigenvalues of the spin projection operator along the $z$-axis. Introducing the new coordinate $x=\xi r$, Eq.~\eqref{4.3} becomes
 \begin{equation}
    x^2\frac{\partial^2\eta_{\ell}}{\partial x^2}
    +x\frac{\partial\eta_{\ell}}{\partial x}
    +\left(x^2-\nu^2\right)\eta_{\ell}=0,
\end{equation}which  corresponds to the Bessel differential equation~\cite{abramowitz1965handbook}. The general radial solution for each spinor component is therefore
\begin{equation}
     \eta_{\ell}(\xi r)=A J_{|\nu|}(\xi r)+B Y_{|\nu|}(\xi r),
    \label{genSolution}
\end{equation}where  $J_{|\nu|}$ and $Y_{|\nu|}$ are Bessel functions of the first and second  kind, respectively. Once we have this general solution in hand, we can explore different boundary conditions to configure the positions of the hard walls where the fermions must be located.

\section{nonlocal effects on fermions confined in a cylinder}\label{cylinder}

The first case we consider here is the confinement of fermions in a cylindrical box of radius $r=r_0$. In order to find its bound-state spectrum, one must set the wave function to be regular at the origin and therefore the coefficient $B$ in Eq.~\eqref{genSolution} has to be zero, since in this case the Bessel function of the second kind $Y_{|\nu|}(\xi r)$ diverges at $r=0$. The remaining solution satisfies the hard-wall boundary condition
\begin{equation}
     \eta_{\ell}(\xi r_0)=0,
    \label{cylindBC}
\end{equation}
which enforces the vanishing of the wave function at the cylindrical wall.

\begin{figure}[t!]
    \centering
    \includegraphics[width=8.5cm]{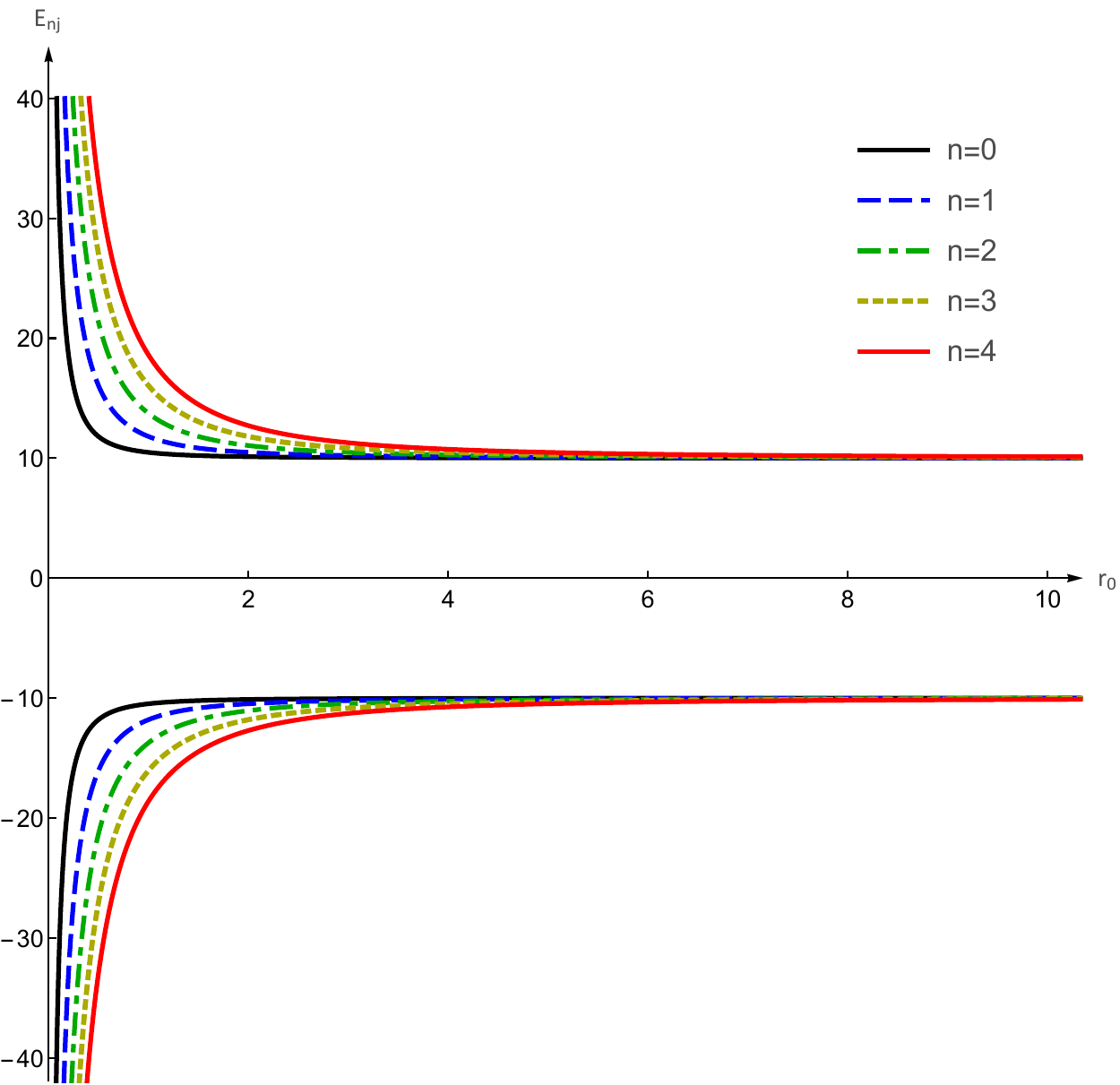}
    \caption{Energy levels of fermions confined in the cylinder as aa functions of its radius, given by Eq. \eqref{energylevels} for different values of $n$. Here, we use $m=\lambda=10^{-3}$, $\Lambda=10^{4}\lambda$, $\kappa=0$, $j=0$ and $\sigma_z=1$.}
    \label{fig1}
\end{figure}

In partcular, we are interested in the regime where a fermion confined within a cylinder behaves as a quantum dot~\cite{bakke2013noninertial, azevedo1998topological}, namely $\xi r_0\gg1$. In this limit~\cite{abramowitz1965handbook}, the Bessel function of the first kind has the asymptotic form
\begin{equation}
     J_{|\nu|}(\xi r_0)\approx
    \sqrt{\frac{2}{\pi\xi r_0}}
    \cos\!\left(\xi r_0-\frac{\pi}{2}|\nu|-\frac{\pi}{4}\right)
    \label{dotquantum}
\end{equation}
and the boundary condition~\eqref{cylindBC}  provides the approximate quantization condition
\begin{equation}
 \xi r_0=\frac{\pi}{2}
    \left[(2n+1)+|\nu|+\frac{1}{2}\right],
    \label{RC1}
\end{equation} where $n$ is the radial cylindrical mode. Using Eqs.~\eqref{xiDef} and~\eqref{nuDef}, the corresponding energy levels are
{\small \begin{equation}
 E_{n,j}=\pm\sqrt{
    k^2\!+\!M_{\rm eff}^{2}\!
    +\!\left(\frac{\pi}{2r_0}\right)^2
    \!\left[(2n+1)+\left|j-\frac{\sigma_z}{2}\right|\!+\!\frac{1}{2}\right]^2}.
    \label{energylevels}
\end{equation}}%
This equation represents the energy levels for fermions confined in the region $0\leq r\leq r_0$ by a cylindrical hard-wall potential. Its behavior is depicted in Fig. \eqref{fig1}, where one can observe that for sufficiently large $r_0$ values 
the energy spectrum of the fermion field put aside its discrete, quantum character, approaching a constant which depends on the nonlocal parameter.

We have also to point out that the result in Eq.~\eqref{energylevels} emerges from the confinement of the fermion in a physical region bounded by a cylindrical shell. In this way, the bounded states result from the system's geometry, which acts as an effective hard-wall confining potential. The energy levels explicitly depends on the confinement radius $r_0$, but nonlocal effects, on the other hand, are only encoded in the effective fermion mass $M_{\rm eff}$ defined in Eq.~\eqref{efffM},  thereby shifting the mass contribution to the spectrum and, consequently, the energy separation gap. This result is analogous to a quantum dot described by a hard-wall potential in a condensed-matter system ~\cite{rosas2000electron, solimany1995electron, aquino2003aharonov, bakke2011analogue}. 

For completeness, let us make the local limit explicit. By defining the positive local energy as
{\small\begin{equation}
     E^{(0)}_{n,j}=\sqrt{
    k^2+m^2
    +\left(\frac{\pi}{2r_0}\right)^2
    \left[(2n+1)+\left|j-\frac{\sigma_z}{2}\right|+\frac{1}{2}\right]^2}.
    \label{local_cylinder_energy}
\end{equation}}%
and using Eq.~\eqref{meff_expansion}, one finds
\begin{eqnarray}
  E_{n,j}=&\pm&\left\{
    E^{(0)}_{n,j}
    +\frac{m^3}{E^{(0)}_{n,j}\Lambda}+\right.\\[4pt]
 \nonumber   &~&\left.+\frac{1}{\Lambda^2}
    \left[
    \frac{2m^4}{E^{(0)}_{n,j}}
    -\frac{m^6}{2\left(E^{(0)}_{n,j}\right)^3}
    \right]
    \!+\!\mathcal{O}(\Lambda^{-3})\right\}\!.~~~~
    \label{cylinder_large_lambda}
\end{eqnarray}
In this way, the leading nonlocal correction is of order $\Lambda^{-1}$ and the standard hard-wall spectrum is recovered as $\Lambda\to\infty$.
\section{nonlocal effects on a quantum ring}\label{ring}
\begin{figure}[t!]
    \centering
    \includegraphics[width=9cm]{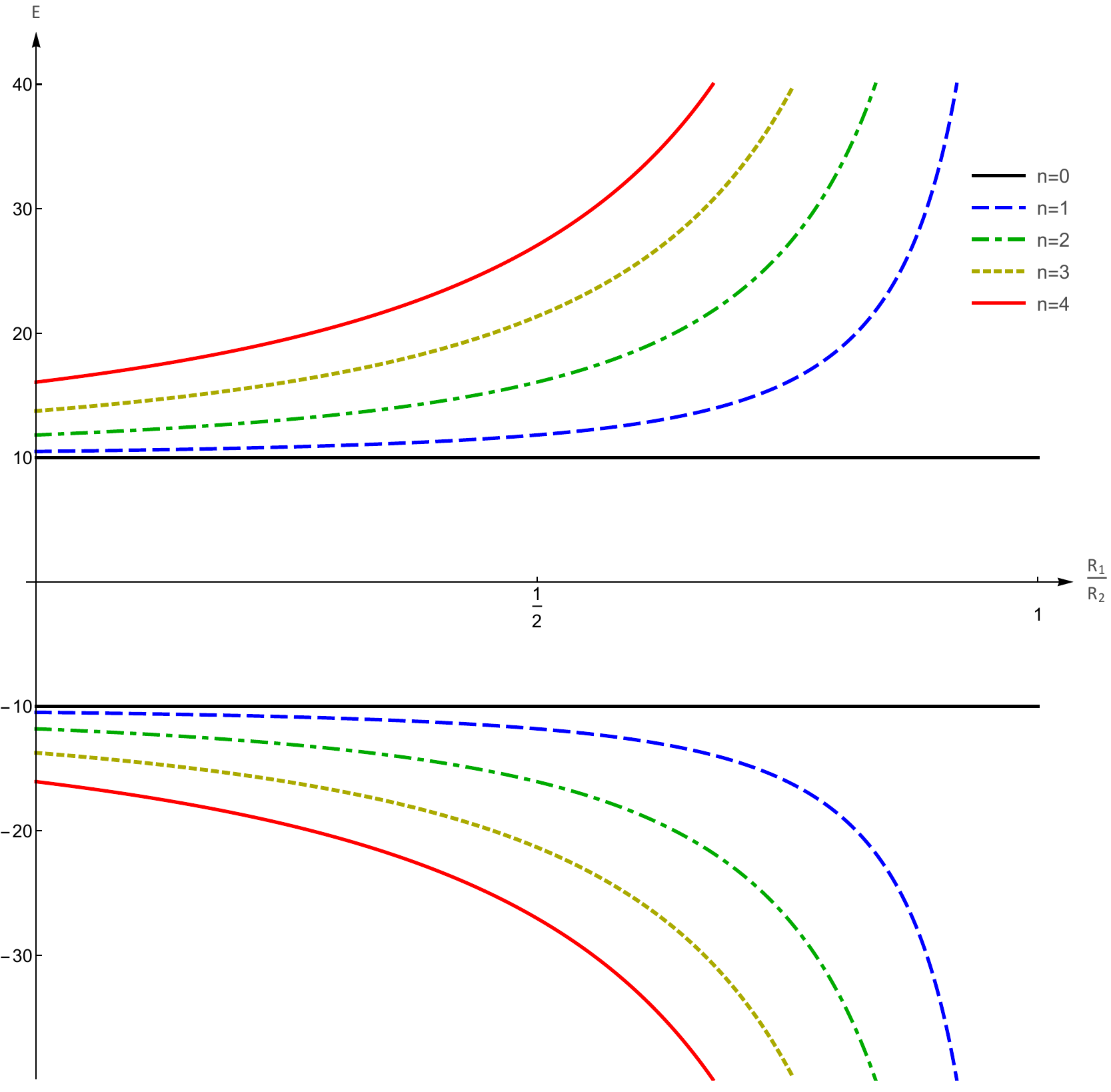}\\
    (a)\\
    \includegraphics[width=9cm]{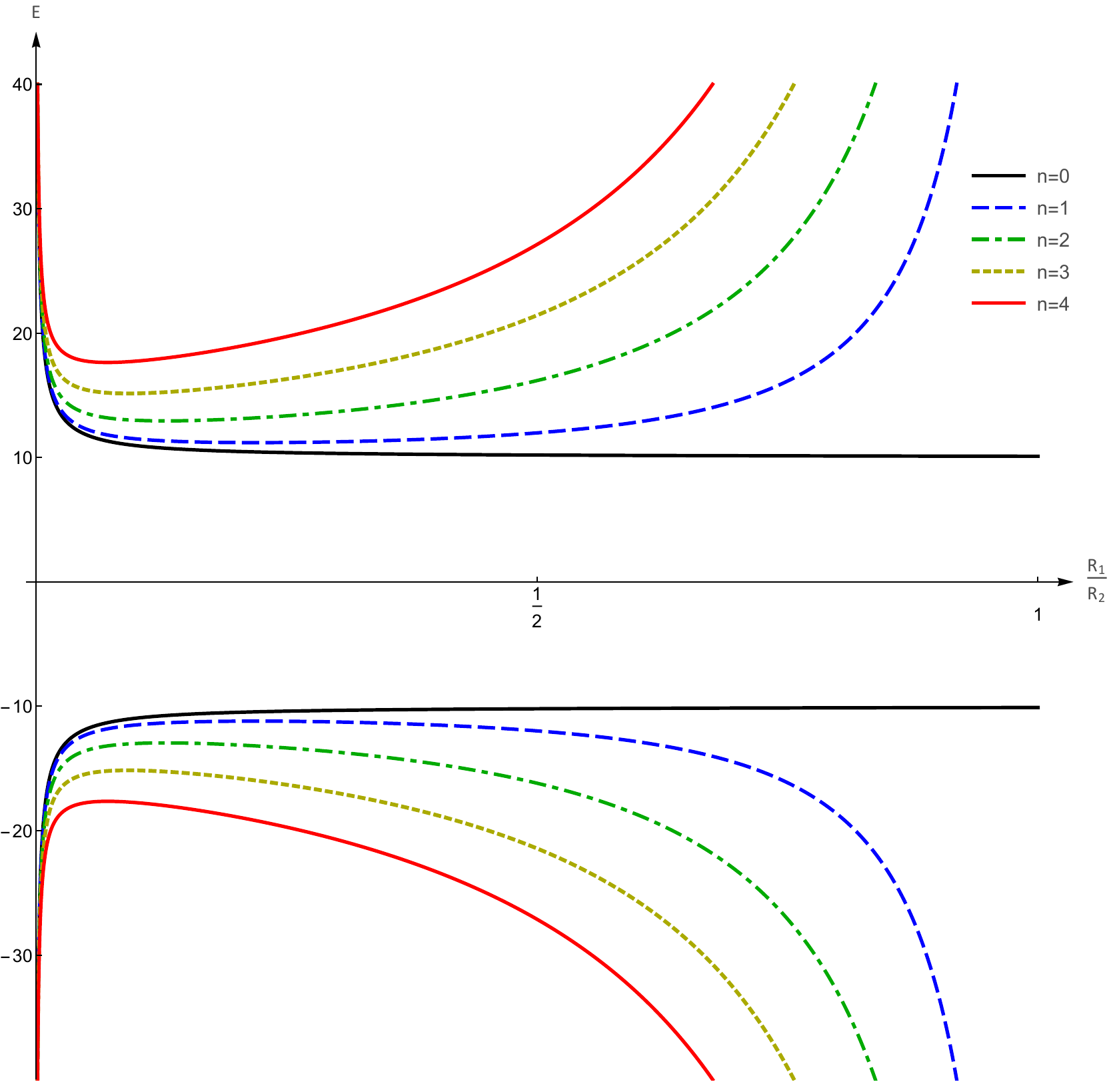}\\
    (b)
    \caption{Energy levels for fermions confined in a ring with inner and outer radii given by $R_1$ and $R_2$, respectively, given by Eq. \eqref{ringEnergy} for different values of $n$. Here, we use $m=\lambda=10^{-3}$, $\Lambda=10^{4}\lambda$, $\kappa=0$, $\sigma_z=1$ and fix $R_2=1$. In (a) we consider setups with $j=0$ and in (b) we consider scenarios with $j=2$.}
    \label{fig2}
\end{figure}

The second system addressed here deal with a fermion confined in a quantum ring by a hard-wall potential. We consider the physical annular region between two cylindrical shells,   $R_1\leq r\leq R_2$. The radial wave function~\eqref{genSolution} in this case must satisfy the boundary conditions $\eta_{\ell}(\xi R_1)=\eta_{\ell}(\xi R_2)=0$, which implies that we have to solve the system
\begin{subequations}
     \begin{eqnarray}
         A J_{|\nu|}(\xi R_1)+B Y_{|\nu|}(\xi R_1)&=& 0,\\
         A J_{|\nu|}(\xi R_2)+B Y_{|\nu|}(\xi R_2)&=& 0,
    \end{eqnarray}
    \end{subequations}
or, in matrix form,
 \begin{equation}
    {\bf \begin{bmatrix}
        J_{|\nu|}(\xi R_1) & Y_{|\nu|}(\xi R_1)\\
        J_{|\nu|}(\xi R_2) & Y_{|\nu|}(\xi R_2)
    \end{bmatrix}
    \begin{bmatrix}A\\B\end{bmatrix}=0.
    \label{BesselMatrix}
}
\end{equation}%
For nontrivial solutions, the determinant of the Bessel-function matrix must vanish. Thus,
\begin{equation}
     J_{|\nu|}(\xi R_1)Y_{|\nu|}(\xi R_2)
    -J_{|\nu|}(\xi R_2)Y_{|\nu|}(\xi R_1)=0.
    \label{nonTrivialSol}
\end{equation}
In order to find an analytical expression for the  energy eigenvalues, we  consider $\xi R_1\gg1$ and $\xi R_2\gg1$. In this case, for fixed $\nu$,  Hankel's asymptotic expansion~\cite{abramowitz1965handbook} provides the following approximations: 
{\small \begin{subequations}
 \begin{eqnarray}
\nonumber
J_{|\nu|}(\xi R_{\ell})&\approx & \sqrt{\frac{2}{\pi\xi R_{\ell}}}
\left[\cos\!\left(\xi R_{\ell}-\nu\frac{\pi}{2}-\frac{\pi}{4}\right)-\right.\\[3pt]
&& \left.\hspace{0.8cm}-\frac{4\nu^2-1}{8\xi R_{\ell}}
\sin\!\left(\xi R_{\ell}-\nu\frac{\pi}{2}-\frac{\pi}{4}\right)\right],~~~~~\\[5pt]
 \nonumber
Y_{|\nu|}(\xi R_{\ell})& \approx & \sqrt{\frac{2}{\pi\xi R_{\ell}}}
\left[\sin\!\left(\xi R_{\ell}-\nu\frac{\pi}{2}-\frac{\pi}{4}\right)-\right.\\[3pt]
&& \left.\hspace{0.8cm}-\frac{4\nu^2-1}{8\xi R_{\ell}}
\cos\!\left(\xi R_{\ell}-\nu\frac{\pi}{2}-\frac{\pi}{4}\right)\right].
\end{eqnarray}\end{subequations}}%
Substituting these expressions into Eq.~\eqref{nonTrivialSol} and defining $\alpha=(4\nu^2-1)/(8\xi R_1R_2)$, we find
\begin{equation}
     \tan\!\left[\xi(R_2-R_1)\right]
    =\frac{\alpha(R_2-R_1)}{1+\alpha^2R_1R_2}.
\end{equation}
Since $\xi R_1\gg1$ and $\xi R_2\gg1$, the right-hand side of the expression above is small and therefore one can write $\xi(R_2-R_1)=n\pi+\epsilon$, with $\epsilon\ll1$  and $n=0,1,2,\cdots$, which lead us to the expression 
\begin{equation}
    \xi^2\approx
    \frac{n^2\pi^2}{(R_2-R_1)^2}
    +\frac{4\nu^2-1}{4R_1R_2}.
    \label{ringXi}
\end{equation} 
Substituting this relation into Eq.~\eqref{xiDef} and manipulating it together with Eq.~\eqref{nuDef}, the analytical energy eigenvalues for nonlocal fermions confined in the quantum ring becomes
{\small \begin{equation}
     E_{n,j}=\pm\sqrt{
    \frac{n^2\pi^2}{(R_2\!-\!R_1)^2}
    \!+\!\frac{1}{R_1R_2}\left[\left(j\!-\!\frac{\sigma_z}{2}\right)^{\!2}\!-\!\frac{1}{4}\right]
    \!+\!k^2\!+\!M_{\rm eff}^{2}}.
    \label{ringEnergy}
\end{equation}}%
 As in the cylindrical case, one can observe here that nonlocality modifies the energy spectrum through the effective mass, while the geometrical hard-wall quantization terms retain their local form. Moreover, defining the positive local ring energy as the limit $E^{(0)}_{n,j}=\lim_{\Lambda\to\infty}E_{n,j}$, we have
{\small\begin{equation}
 E^{(0)}_{n,j}=\sqrt{
    \frac{n^2\pi^2}{(R_2-R_1)^2}
    +\frac{1}{R_1R_2}\left[\left(j-\frac{\sigma_z}{2}\right)^2\!-\!\frac{1}{4}\right]
    \!+\!k^2\!+\!m^2}.
\label{local_ring_energy}
\end{equation}}%
and the large-$\Lambda$ expansion becomes
\begin{eqnarray}
E_{n,j}=&\pm&\!\!\left\{
E^{(0)}_{n,j;}
+\frac{m^3}{E^{(0)}_{n,j}\Lambda}+\right.\\[4pt]
\nonumber &~&\left.+\frac{1}{\Lambda^2}
\left[
\frac{2m^4}{E^{(0)}_{n,j}}
-\frac{m^6}{2\left(E^{(0)}_{n,j}\right)^3}
\right]
+\!\mathcal{O}(\Lambda^{-3})\!\right\}\!.~~~~~~
\label{ring_large_lambda}
\end{eqnarray}
Another interesting fact about these models is that non-local effects also do not affect the presence of degenerate states in the energy spectrum. In fact, in both the problem addressed here and the previous one, the cylinder problem, the states where $\{(j,\sigma_z)\}=\{(0,\pm1),(1,1)\}$ are degenerate. In the previous case, in particular, this degeneracy does not qualitatively affect the energy states of confined fermions. In the present case, however, we observe that such degeneracy affects the limit where $R_1\to 0$. In this limit, the non-degenerate states, which occur for $j\geq 2$ and for $(j,\sigma_z)=(1,-1)$, show a divergence, as expected, since in this setup the quantum ring resembles a cylinder of radius $R_2$. The degenerate states mentioned above, in turn, exhibit distinct behavior, approaching a finite value as $R_1\to0$. These distinct behaviors are shown in Figs. (\ref{fig2}a) and (\ref{fig2}b), where we illustrate the energy levels \eqref{ringEnergy} for different values of $n$. It is also important to note that, in the limit for massless fermions, when $m=0$, the Lambert function $W_{0}\left(-\frac{m}{\Lambda}\right)$ also goes to zero, so that for $n=0$ and $j=0$ in equation~\eqref{ringEnergy} we obtain zero modes. Zero modes are intimately related to topological insulators. This is an interesting result because it suggests that a description of graphene with non-local effects could lead to a new kind of topological insulator, where the non-local parameter could be related to the topological charge. 

\section{Concluding remarks}\label{conclusions}

In this work, we investigated nonlocal effects on the energy spectrum of spin-1/2 fermions confined by hard-wall potentials. Our main result is that the nonlocal modification of the
fermionic dynamics can be effectively incorporated into the mass sector
of the theory. On the physical branch continuously connected to the local
Dirac theory, the nonlocality scale modifies the fermion mass in a
well-defined manner, while preserving a smooth local limit as
$\Lambda\rightarrow\infty$. In particular, the large-$\Lambda$ expansion
shows that the leading nonlocal correction appears at order
$\Lambda^{-1}$.

For both confinement geometries considered here, namely the cylindrical
hard-wall potential and the quantum ring, the geometrical quantization
conditions retain the same structure as in the corresponding local
problem. The nonlocal contribution enters through the effective mass and
therefore produces a systematic shift of the energy levels without
altering the geometrical dependence on the confinement parameters and
quantum numbers. As a consequence, the standard hard-wall spectra are
continuously recovered in the local limit.

Another relevant result is that, within the present model, nonlocality
does not lift the degeneracy of states that are already degenerate in the
local theory. Since the nonlocal correction is independent of the radial,
angular, and spin quantum numbers, states sharing the same geometrical
quantization contribution receive the same energy shift. Likewise, the
symmetry between the positive- and negative-energy branches is preserved,
although their separation is modified by the nonlocal effective mass.

These results show that hard-wall confinement and nonlocality affect
distinct sectors of the spectrum: the former controls the geometrical
mode quantization, whereas the latter modifies the fermionic mass
contribution. This separation makes it possible to identify the genuine
spectral effects associated with nonlocality in confined fermionic
systems and provides a useful framework for further investigations in
the presence of external fields (for example, electromagnetic fields), topological defects or curved
backgrounds with cylindrical symmetry, as graphene wormhole models \cite{garcia2020graphene,garcia2025rotation}.
\acknowledgments{This work was par\-tial\-ly sup\-por\-ted by PRO\-NEX/FA\-PESQ-PB, CNPq and CA\-PES (PRO\-CAD). G.Q. Garcia would like to thank Fapesq-PB for financial support (Grant BLD-ADT-A2377/2024). P.J. Porfírio would like to thank the Brazilian agency CNPq for financial support (PQ–C
grant, process No. 307011/2026-7). E. Brito  would like to thank  FAPESB (Fundação de Amparo à Pesquisa do Estado da Bahia) for financial support (Grant No APP0041/2023 and PPP0006/2024). D.C. Moreira acknowledges partial support from UEPB and from the National Council for Scientific and Technological Development, CNPq, under grant 310838/2026-6.}

\bibliography{references.bib}  

\end{document}